\documentclass[UTF-8,preprint,showpacs,preprintnumbers,superscriptaddress]{revtex4-1}
\usepackage{multirow}
\usepackage{diagbox}
\usepackage{booktabs}
\usepackage{array}
\usepackage{natbib}
\usepackage{amsmath,amssymb,graphicx,bm,subfigure,float,siunitx,color}
\usepackage{graphicx}
\usepackage[utf8]{inputenc}
\usepackage[figuresright]{rotating}
\usepackage{color}
\usepackage{epstopdf}
\usepackage{mathrsfs}
\usepackage{subcaption}
\usepackage{caption}
\begin{document}

\title{Nonsingular hairy black holes by gravitational decoupling}

\author{Yaobin Hua}
\affiliation{College of Physics Science and Technology, Hebei University, Baoding 071002, China}
\author{Rongjia Yang \footnote{Corresponding author}}
\email{yangrongjia@tsinghua.org.cn}
\affiliation{College of Physics Science and Technology, Hebei University, Baoding 071002, China}
\affiliation{Hebei Key Lab of Optic-Electronic Information and Materials, Hebei University, Baoding 071002, China}
\affiliation{National-Local Joint Engineering Laboratory of New Energy Photoelectric Devices, Hebei University, Baoding 071002, China}
\affiliation{Key Laboratory of High-pricision Computation and Application of Quantum Field Theory of Hebei Province, Hebei University, Baoding 071002, China}

\begin{abstract}
Using gravitational decoupling under the requirements of a well-defined event horizon and the source matter satisfying the weak energy condition, we construct nonsingular hairy black holes with spherical or axial symmetry. These solutions emerge from a deformation of the Minkowski vacuum, bridging the novel hairy geometries and the classical Schwarzschild and Kerr solutions at the maximum deformation in their respective sectors.
\end{abstract}

\maketitle

\section{introduction}

The avenues for circumventing the no-hair conjecture have been extensively explored \cite{Martinez:2004nb,Sotiriou:2011dz,babichev2014dressing,Antoniou:2017acq,Antoniou:2017hxj,Sotiriou:2013qea}. One promising approach is to introduce an additional source of potential fundamental origin—often a scalar field \cite{Herdeiro:2015waa}—into the static vacuum of General Relativity (GR). A key motivation is to eradicate the singularities that GR predicts as the endpoint of gravitational collapse. Notwithstanding the Cosmic Censorship Conjecture's (CCC) assertion that these singularities are hidden behind horizons \cite{Penrose:1969pc,hawking2023large}, their theoretical existence points to a fundamental limitation of GR.

In the recent pursuit of singularity-free hairy black holes (BHs), models based on non-linear electrodynamics have emerged as a relatively straightforward framework for sourcing regular geometries \cite{Salazar:1987ap,Ayon-Beato:1998hmi,Bronnikov:2000vy,Dymnikova:2004zc,Balart:2014cga,Toshmatov:2014nya,Fan:2016hvf}. A significant challenge is that such classically singularity-free solutions typically possess a Cauchy horizon. As a null hypersurface, it compromises deterministic predictability \cite{Poisson:1989zz,Poisson:1990eh} and introduces numerous theoretical pathologies \cite{Bonanno:2020fgp,carballo2022regular,Franzin:2022wai,Bonanno:2022jjp,Ovalle:2023ref}. A promising strategy to evade these difficulties is to adopt a maximally general description of matter, which provides a flexible scenario governed by minimal assumptions.

In this work, we adopt this approach by filling the Schwarzschild vacuum with a generic static, spherically symmetric source $\theta_{\mu\nu}$, a ``tensor vacuum", via the gravitational decoupling (GD) formalism \cite{Ovalle:2017fgl,Ovalle:2018gic}—a powerful technique for generating hairy BHs in spherical \cite{Ovalle:2018umz,Ovalle:2020kpd} and axial symmetry \cite{Contreras:2021yxe,Islam:2021dyk,daRocha:2020gee,Ovalle:2021jzf,afrin2021parameter,Ramos:2021jta,Meert:2021khi,Mahapatra:2022xea,Cavalcanti:2022cga,Omwoyo:2021uah,Avalos:2023jeh,Avalos:2023ywb,Hua:2025qwu}. 
The principal advantage of this scheme lies in its ability to enforce minimal physical requirements while maintaining asymptotic flatness. We therefore seek regular BH solutions that satisfy the Weak Energy Condition (WEC) for both static and rotating configurations. 
Then, we have successfully constructed such a non-asymptotically flat, regular BH and systematically elucidated the influence of the relevant parameters on its properties.

This paper is structured as follows. Section \MakeUppercase{\romannumeral2} presents the fundamentals of the GD scheme and details the decoupling of two gravitational sources under spherical symmetry. Building on this, Section \MakeUppercase{\romannumeral3} constructs spherically symmetric, regular hairy BHs that adhere to the WEC. The axially symmetric counterpart of this regular hairy BH is derived in Section \MakeUppercase{\romannumeral4}.  Finally, our conclusions are summarized in Section \MakeUppercase{\romannumeral5}.

In this paper, we use units with $c=1$ and $\kappa=8\,\pi\,G$, where $G$ is Newton's constant, and the adopted signature is $(-,+,+,+)$.

\section{gravitational decoupling}

We begin by briefly outlining the GD formalism for spherically symmetric gravitational systems; for a comprehensive treatment, we refer the reader to \cite{Ovalle:2020kpd}. Our starting point is the Einstein-Hilbert action
\begin{align}
\label{EHA}
\mathcal{S}=\int \bigg[\frac{R}{2\kappa}+\mathcal{L}_{M}+\mathcal{L}_{\Theta} \bigg]\sqrt{-g}\ d^{4}x,
\end{align}
where $R$ is the Ricci scalar, $\mathcal{L}_{\rm M}$ corresponds to the standard matter fields, and
$\mathcal{L}_{\Theta}$ is the second Lagrangian density, which can describe matter or be related to new gravitational sectors beyond GR. 
For these two sources, the energy-momentum tensor is generally defined, respectively, as
\begin{align}
T_{\mu\nu}&=-\frac{2}{\sqrt{-g}}\frac{\delta\, \big(\sqrt{-g}\,\mathcal{L}_{\rm M}\big)}{\delta \,g^{\mu\nu}}=g_{\mu\nu}\,\mathcal{L}_{\rm M}-2\,\frac{\delta\,\mathcal{L}_{\rm M}}{\delta \,g^{\mu\nu}}, \label{Tuv}   \\
\theta_{\mu\nu}&=-\frac{2}{\sqrt{-g}}\frac{\delta\, \big(\sqrt{-g}\,\mathcal{L}_{\Theta}\big)}{\delta \,g^{\mu\nu}}=g_{\mu\nu}\,\mathcal{L}_{\Theta}-2\,\frac{\delta\,\mathcal{L}_{\Theta}}{\delta \,g^{\mu\nu}}.\label{thetauv}
\end{align}
The Einstein field equations are derived from the action \eqref{EHA} by adopting the standard procedure
\begin{align}
\label{Guv}
G_{\mu\nu}=R_{\mu\nu}-\frac{1}{2}R\,g_{\mu\nu}=\kappa \,\tilde{T}_{\mu\nu}=\kappa \,\big(T_{\mu\nu}+\theta_{\mu\nu}\big).
\end{align}
where $\tilde{T}_{\mu\nu}$ represents the total energy-momentum tensor, namely: $\tilde{T}_{\mu\nu}=T_{\mu\nu}+\theta_{\mu\nu}$.
Moreover, as a fundamental consequence of the second Bianchi identity, the covariant divergence of the Einstein tensor vanishes, leading to the covariant conservation equation
\begin{align}
\label{nablaGuv}
\nabla_{\mu} \,G^{\mu\nu}=\kappa\,\nabla_{\mu} \, \tilde{T}^{\mu\nu}=\kappa \, \nabla_{\mu}\,\big(T^{\mu\nu}+\theta^{\mu\nu}\big)=0.
\end{align}
The metric for a static and spherically symmetric spacetime may be described by
\begin{align}
\label{1g}
ds^{2}=-e^{A(r)}dt^{2}+e^{B(r)}dr^{2}+r^{2}\Big(d\theta^{2}+\sin^{2}\theta d\phi^{2}\Big),
\end{align}
where $A=A(r)$ and $B=B(r)$ depend solely on the radial coordinate $r$. 
Then, from the Einstein field equations \eqref{Guv}, we obtain 
\begin{align}
G_{0}^{\ 0}&=\kappa \,\tilde{T}_{0}^{\ 0}=\kappa \,\big(T_{0}^{\ 0}+\theta_{0}^{\ 0}\big)=-\frac{1}{r^{2}}+e^{-B}\Bigg(\frac{1}{r^{2}}-\frac{B'}{r}\Bigg),   \label{G00} \\
G_{1}^{\ 1}&=\kappa \,\tilde{T}_{1}^{\ 1}=\kappa \,\big(T_{1}^{\ 1}+\theta_{1}^{\ 1}\big)=-\frac{1}{r^{2}}+e^{-B}\Bigg(\frac{1}{r^{2}}+\frac{A'}{r}\Bigg),    \label{G11} \\
G_{2}^{\ 2}&=\kappa \,\tilde{T}_{2}^{\ 2}=\kappa \,\big(T_{2}^{\ 2}+\theta_{2}^{\ 2}\big)=\frac{e^{-B}}{4}\Bigg(2A''+A'^{2}-A'B'+2\,\frac{A'-B'}{r}\Bigg),   \label{G22} \\
G_{3}^{\ 3}&=\kappa \,\tilde{T}_{3}^{\ 3}=\kappa \,\big(T_{3}^{\ 3}+\theta_{3}^{\ 3}\big)=\frac{e^{-B}}{4}\Bigg(2A''+A'^{2}-A'B'+2\,\frac{A'-B'}{r}\Bigg),   \label{G33}
\end{align}
where $f'\equiv\partial_{r}$. Due to the spherical symmetry of the metric, it is easy to see that $G_{2}^{\ 2}=G_{3}^{\ 3}$.
Next, we identify, respectively, an effective energy density $\tilde{\epsilon}$, an effective radial pressure $\tilde{p_{r}}$, and an effective tangential pressure $\tilde{p_{t}}$ as
\begin{align}
\tilde{\epsilon}&\equiv \epsilon+\mathcal{E}=-T_{0}^{\ 0}-\theta_{0}^{\ 0},                \label{epsilon}  \\
\tilde{p}_{r}&\equiv p_{r}+\mathcal{P}_{r}=T_{1}^{\ 1}+\theta_{1}^{\ 1,}                   \label{pr}       \\
\tilde{p}_{t}&\equiv p_{\theta}+\mathcal{P}_{\theta}=T_{2}^{\ 2}+\theta_{2}^{\ 2},    \label{pt}
\end{align}
where $\epsilon,~ p_{r},~ p_{\theta}$ are related to $T_{\mu}^{\ \nu}$ and $\mathcal{E},~ \mathcal{P}_{r},~ \mathcal{P}_{\theta}$ are related to $\theta_{\mu}^{\ \nu}$. Then, we have
\begin{align}
T_{\mu}^{\ \nu}={\rm diag}\big[-\epsilon,~ p_{r},~ p_{\theta},~ p_{\theta}\big];~\,\theta_{\mu}^{\ \nu}={\rm diag}\big[-\mathcal{E},~ \mathcal{P}_{r},~ \mathcal{P}_{\theta},~ \mathcal{P}_{\theta}\big]. \label{diagTanddiagtheta} 
\end{align}
In general, Eqs. \eqref{G00}-\eqref{G22} describe an anisotropic fluid, and $\Pi\equiv\tilde{p}_{t}-\tilde{p}_{r}\neq0$.

Let us denote the solution to Eq. \eqref{Guv} generated by the seed source $T_{\mu\nu}$ alone as the ``seed solution", which has the metric
\begin{align}
\label{2g}
ds^{2}=-e^{D(r)}dt^{2}+e^{E(r)}dr^{2}+r^{2}\Big(d\theta^{2}+\sin^{2}\theta d\phi^{2}\Big), 
\end{align}
where
\begin{align}
\label{eFm}
e^{-E(r)}\equiv1+\frac{\kappa}{r}\int_{0}^{r} x^{2}\,T_{0}^{\ 0}\,dx=1-\frac{2\,m(r)}{r},
\end{align}
which is the standard expression in GR for $m(r)$ being the Misner-Sharp mass $m$.

The GD of the metric \eqref{2g} by the introduction of the source $\theta_{\mu\nu}$ yields
\begin{align}
D(r)&\longrightarrow A(r)=D(r)+\alpha\,g(r),                      \label{jiantou1} \\
e^{-E(r)}&\longrightarrow e^{-B(r)}=e^{-E(r)}+\alpha\,f(r),       \label{jiantou2}
\end{align}
where $f$ and $g$ denote the geometric deformations for the temporal and radial metric components, respectively, controlled by the parameter $\alpha$. We emphasize that the ansatz in Eqs.\eqref{jiantou1}-\eqref{jiantou2} describes a physical deformation of the spacetime, not a coordinate transformation. Consequently, considering Eqs.\eqref{jiantou1}-\eqref{jiantou2}, the Einstein field Eqs.\eqref{G00}-\eqref{G22} separate into two distinct parts:

The first set is the standard Einstein system for the seed metric \eqref{2g} with source $T_{\mu\nu}$
\begin{align}
\kappa \, \epsilon&=-\kappa \, T_{0}^{\ 0}=\frac{1}{r^{2}}-e^{-E}\Bigg(\frac{1}{r^{2}}-\frac{E'}{r}\Bigg),                                        \label{GD11} \\
\kappa \, p_{r}&=\kappa \, T_{1}^{\ 1}=-\frac{1}{r^{2}}+e^{-E}\Bigg(\frac{1}{r^{2}}+\frac{D'}{r}\Bigg),                                           \label{GD12} \\
\kappa \, p_{\theta}&=\kappa \,T_{2}^{\ 2}=\frac{e^{-E}}{4}\Bigg(2D''+D'^{2}-D'E'+2\,\frac{D'-E'}{r}\Bigg).                      \label{GD13} 
\end{align}
The second set, which includes the source $\theta_{\mu\nu}$, is
\begin{align}
\kappa \, \mathcal{E}&=-\kappa \, \theta_{0}^{\ 0}=-\frac{\alpha\,f}{r^{2}}--\frac{\alpha\,f'}{r},             \label{GD21} \\
\kappa \, \mathcal{P}_{r}&=\kappa \, \theta_{1}^{\ 1}=\alpha\,f\Bigg(\frac{1}{r^{2}}+\frac{A'}{r}\Bigg)+\alpha\,X_{1},             \label{GD22} \\
\kappa \, \mathcal{P}_{\theta}&=\kappa \,\theta_{2}^{\ 2}=\frac{\alpha\,f}{4}\Bigg(2A''+A'^{2}+2\,\frac{A'}{r}\Bigg)+\frac{\alpha\,f'}{4}\Bigg(A'+\frac{2}{r}\Bigg)+\alpha\,X_{2}, \label{GD23} 
\end{align}
where
\begin{align}
X_{1} &=\frac{e^{-E}\,g'}{r},                                                  \label{X1} \\
4X_{2} &=e^{-E}\bigg(2\,g''+\alpha\,g'^{2}+\frac{2\,g'}{r}+2\,g'D'-E'g' \bigg).       \label{X2}
\end{align}
It is obvious that the effective source $\theta_{\mu\nu}$ vanishes when the geometric deformations are switched off $(f=g=0)$.

Next, through Eq.\eqref{nablaGuv}, we find that the conservation equation is a linear combination of Eqs.\eqref{G00}-\eqref{G22}, namely,
\begin{align}
\label{shouheng1}
\big(\tilde{T}_{1}^{\ 1}\big)'-\frac{A'}{2}\big(\tilde{T}_{0}^{\ 0}-\tilde{T}_{1}^{\ 1}\big)-\frac{2}{r}\big(\tilde{T}_{2}^{\ 2}-\tilde{T}_{1}^{\ 1}\big)=0.
\end{align}
Since there are two sources, this formula can be decomposed into
\begin{align}
\label{shouheng2}
\nabla\big(T_{\mu\nu}+\theta_{\mu\nu}\big)=&\big(T_{1}^{\ 1}\big)'-\frac{A'}{2}\big(T_{0}^{\ 0}-T_{1}^{\ 1}\big)-\frac{2}{r}\big(T_{2}^{\ 2}-T_{1}^{\ 1}\big)\nonumber\\
&+\big(\theta_{1}^{\ 1}\big)'-\frac{A'}{2}\big(\theta_{0}^{\ 0}-\theta_{1}^{\ 1}\big)-\frac{2}{r}\big(\theta_{2}^{\ 2}-\theta_{1}^{\ 1}\big)=0.
\end{align}
After substituting Eq.\eqref{jiantou1} into Eq.\eqref{shouheng2}, Eq.\eqref{shouheng2} can be expressed as
\begin{align}
\label{shouheng3}
\bigg[\big(T_{1}^{\ 1}\big)'-\frac{D'}{2}\big(T_{0}^{\ 0}-T_{1}^{\ 1}\big)-\frac{2}{r}\big(T_{2}^{\ 2}-T_{1}^{\ 1}\big)\bigg]-\frac{\alpha \,g'}{2}\big(T_{0}^{\ 0}-T_{1}^{\ 1}\big)& \nonumber \\
+\big(\theta_{1}^{\ 1}\big)'-\frac{A'}{2}\big(\theta_{0}^{\ 0}-\theta_{1}^{\ 1}\big)-\frac{2}{r}\big(\theta_{2}^{\ 2}-\theta_{1}^{\ 1}\big)&=0.
\end{align}
From the conservation equation $\nabla_{\mu}\,T^{\mu\nu}=0$, we have
\begin{align}
\label{shouheng4}
\bigg[\big(T_{1}^{\ 1}\big)'-\frac{D'}{2}\big(T_{0}^{\ 0}-T_{1}^{\ 1}\big)-\frac{2}{r}\big(T_{2}^{\ 2}-T_{1}^{\ 1}\big)\bigg]=0.
\end{align}
Eq. \eqref{shouheng4} is a linear combination of Eqs. \eqref{GD11}-\eqref{GD13}. Then we can obtain
\begin{align}
\label{shouheng5}
\nabla_{\mu}^{({\rm A,B})}\,T^{\ \mu}_{\nu}=\nabla_{\mu}^{({\rm D,E})}\,T^{\ \mu}_{\nu}-\frac{\alpha \,g'}{2}\big(T_{0}^{\ 0}-T_{1}^{\ 1}\big).
\end{align}
where $({\rm A,B})$ is related to the metric \eqref{1g}, and $({\rm D,E})$ is related to the metric \eqref{2g}. Eq. \eqref{shouheng5} can also be rewritten as
\begin{align}
\nabla_{\mu}\,T^{\ \mu}_{\nu}&=-\frac{\alpha \,g'}{2}\big(T_{0}^{\ 0}-T_{1}^{\ 1}\big),  \label{shouheng6} \\
\nabla_{\mu}\,\theta^{\ \mu}_{\nu}&=\frac{\alpha \,g'}{2}\big(T_{0}^{\ 0}-T_{1}^{\ 1}\big),  \label{shouheng7}
\end{align}
where the covariant derivative on the left-hand side of Eqs.\eqref{shouheng6}-\eqref{shouheng7} is related to the metric \eqref{1g}.
Finally, the conservation equation \eqref{nablaGuv} leads to
\begin{align}
\label{shouheng8}
\nabla_{\mu}\,T^{\ \mu}_{\nu}=\frac{\alpha\,g'}{2}\big(\epsilon+p_{r}\big)=-\nabla_{\mu}\,\theta^{\ \mu}_{\nu}.
\end{align}
Eq. \eqref{shouheng8} signifies an energy-momentum exchange between the two gravitational sectors described by Eqs. \eqref{GD11}-\eqref{GD13} and Eqs. \eqref{GD21}-\eqref{GD23}, respectively. In particular, this transfer vanishes when the interaction is purely gravitational, such as for a vanishing temporal deformation $(g = 0)$ or in a Kerr-Schild spacetime $(\epsilon+p_{r}=0)$. It is essential to recognize that this solution is exact and inherently non-perturbative with respect to the deformations $f$ and $g$ \cite{Ovalle:2020fuo}.

\section{Hairy black holes}
We proceed to consider hairy deformations of spherically symmetric BHs in GR, taking the Schwarzschild metric as our starting point.
\begin{align}
\label{seedSsolution}
e^{D}=e^{-E}=1-\frac{2M}{r}.
\end{align}
This metric corresponds to the vacuum solution of Eqs. \eqref{GD11}-\eqref{GD13} and serves as our seed geometry.
And we seek a matter Lagrangian $\mathcal{L}_{\Theta}$, sourcing $\theta_{\mu\nu}$, capable of inducing geometric deformations $f$ and $g$ via Eqs. \eqref{GD11}-\eqref{GD13}, thereby removing the central singularity at $r=0$. 
The system is underdetermined, with three equations for five unknowns $(f,g,\mathcal{E},\mathcal{P}_{r},\mathcal{P}_{\theta})$, allowing for the imposition of supplementary conditions.

\subsection{Horizon structure}

A well-defined horizon structure requires the equivalence of metric functions, $e^{A(r_{\rm h})}=e^{-B(r_{\rm h})}=0$. This ensures the coincidence of the Killing horizon $(e^{A}=0)$ and causal horizon $(e^{-B}=0)$ at the radius $r=r_{\rm h}$. For this feature, we have a sufficient condition $e^{A(r_{\rm h})}=e^{-B(r_{\rm h})}=0$ which for this horizon structure is physically equivalent to the equation of state $-\tilde{\epsilon}=\tilde{p_{r}}$ for the source, as derived from the Einstein equations Eqs. \eqref{G00}-\eqref{G11}.

Obviously, for $T_{\mu\nu}=0$, we will have $-\mathcal{E}=\mathcal{P}_{r}$, which indicates that for a positive energy density, the radial pressure must be negative.
Furthermore, we can rewrite the conservation equation \eqref{shouheng3} as
\begin{align}
\label{shouheng9}
\big(\theta_{1}^{\ 1}\big)'-\frac{A'}{2}\big(\theta_{0}^{\ 0}-\theta_{1}^{\ 1}\big)-\frac{2}{r}\big(\theta_{2}^{\ 2}-\theta_{1}^{\ 1}\big)=\frac{\alpha \,g'}{2}\big(T_{0}^{\ 0}-T_{1}^{\ 1}\big).
\end{align}
Considering $T_{\mu\nu}=0$, we can obtain
\begin{align}
\label{Prprime}
\mathcal{P}_{r}'=\frac{2}{r}\big(\mathcal{P}_{\theta}-\mathcal{P}_{r}\big).
\end{align}
The hydrostatic equilibrium equation ensures the stability of the source $\theta_{\mu\nu}$, precluding its gravitational collapse into the central singularity of the Schwarzschild seed geometry.

From the combination of the seed metric \eqref{seedSsolution}, the deformation relations \eqref{jiantou1}-\eqref{jiantou2}, and the condition $e^{A}=e^{-B}$, we derive
\begin{align}
\label{alphaf}
\alpha\,f=\bigg(1-\frac{2M}{r}\bigg)\bigg(e^{\alpha\,g}-1\bigg).
\end{align}
Then the metric \eqref{1g} can be  expressed in the following alternative form
\begin{align}
\label{3g}
ds^{2}=-\bigg(1-\frac{2M}{r}\bigg)\,h\,dt^{2}+\bigg(1-\frac{2M}{r}\bigg)^{-1}\!h^{-1}\,dr^{2}+r^{2}\Big(d\theta^{2}+\sin^{2}\theta d\phi^{2}\Big),
\end{align}
where $h=e^{\alpha\,g}$ and $g$ has not yet been determined.

\subsection{Weak energy conditions}
Although classical energy conditions are expected to break down in high-curvature environments, they continue to provide a principled approach to exclude manifestly unphysical models \cite{Martin-Moruno:2017exc}. Guided by this, we require the effective source $\theta_{\mu\nu}$ to satisfy the WEC
\begin{align}
\mathcal{E}\ge 0,                                   \label{WEC1}   \\
\mathcal{E}+\mathcal{P}_{r}\ge 0,                   \label{WEC2}   \\
\mathcal{E}+\mathcal{P}_{\theta}\ge 0.              \label{WEC3}   
\end{align}
Obviously, Eq. \eqref{WEC2} has already been automatically satisfied. Then, substituting Eqs. \eqref{GD21} and \eqref{Prprime} into conditions \eqref{WEC1} and \eqref{WEC3} yields the following respective forms
\begin{align}
\kappa\,\mathcal{E}\,r^{2}&=-(r-2M)\,h'-h+1\ge 0,                        \label{WEC4}   \\
2(\mathcal{E}+\mathcal{P}_{\theta})&=-r\,\mathcal{E}'\ge 0.              \label{WEC5}   
\end{align}
Equation (\ref{WEC4}) is identified as a first-order linear differential inequality for the function $h$, and this relation consistently reproduces the seed Schwarzschild metric \eqref{seedSsolution} when the energy density $\mathcal{E}$ is set to zero.
Furthermore, any everywhere-regular solution must satisfy the system (\ref{WEC4})–(\ref{WEC5}), which for a positive energy density $\mathcal{E}$ entails that it monotonically decreases from $r = 0$, namely, $\mathcal{E}' \le 0$.

\subsection{Regular spacetime metric}

An interesting case we find that satisfies the conditions \eqref{WEC4}-\eqref{WEC5} is
\begin{align}
\label{C2kE}
\kappa\,\mathcal{E}=\frac{\omega}{\eta^{2}}e^{-r/l},
\end{align}
where $\omega\ge0$, $l>0$ and $\eta\neq0$. $\eta$ and $l$ have length dimensions. When $l=\eta$, Eq. \eqref{C2kE} reduces to the case discussed in \cite{Ovalle:2023ref}. When $\omega\rightarrow0$, it returns to the seed vacuum case \eqref{seedSsolution}. The derivative of $\mathcal{E}$ is
\begin{align}
\label{C2kED}
\mathcal{E}'=-\frac{\omega}{\eta^{2}\,l\,\kappa\,}\,e^{-r/l}\le0,
\end{align}
which ensures that Eq. \eqref{WEC5} holds true.

\subsubsection{Spherically symmetric case}

Inserting Eq. \eqref{C2kE} into Eq. \eqref{WEC4} and after some algebraic calculations, we find
\begin{align}
\label{2h}
h=\frac{\omega\,e^{-\frac{r}{l}}\left(-r^{2}\,l-2\,r\,l^{2}-2\,l^{3}\right)}{\left(2M-r\right)\,\eta^{2}}+\frac{c_{1}-r}{2M-r},
\end{align}
where $c_{1}$ is a constant. From Eq. \eqref{2h}, we derive the following forms for the metric functions
\begin{align}
\label{7g}
e^{A}=e^{-B}=1-\frac{c_{1}}{r}+\omega\,e^{-\frac{r}{l}}\left(\frac{r\,l}{\eta^{2}}+\frac{2\,l^{2}}{\eta^{2}}+\frac{2\,l^{3}}{r\,\eta^{2}}\right).
\end{align}
Note that the seed mass $M$ does not appear in the metric function \eqref{7g} and the ADM mass is given by $\mathcal{M}=c_1/2$, where $\mathcal{M}$ will be determined below. If $\omega=0$, the Schwarzschild solution will be recovered. While if $r\sim0$, we have
\begin{align}
\label{8g}
e^{A}=e^{-B}\simeq1-\frac{1}{r}\left(c_1-\frac{2\,l^{3}\,\omega}{\eta^{2}}\right)-\frac{r^{2}\,\omega}{3\,\eta^{2}}+\frac{r^{3}\,\omega}{6\,\eta^{2}\,l}+\mathcal{O}(r^{4}).
\end{align}
The disappearance of the central singularity requires $c_{1}=2l^3\omega/\eta^2$, which implies that $\mathcal{M}=l^3\omega/\eta^2$.
Finally, we can rewrite the asymptotically flat metric \eqref{7g} as
\begin{align}
\label{9g}
e^{A}=e^{-B}=1-\frac{2\mathcal{M}}{r}+\frac{e^{-\frac{r\,\omega\,l^{2}}{\mathcal{M}\,\eta^{2}}}}{r\,\mathcal{M}}\left(\frac{r^{2}\,l^{4}\,\omega^{2}}{\eta^{4}}+\frac{2\,r\,\mathcal{M}\,l^{2}\,\omega}{\eta^{2}}+2\,\mathcal{M}^{2}\right).
\end{align}
After we impose the regularity condition, we find that for $\omega\rightarrow0$ Eq. \eqref{9g} reduces to the Minkowski spacetime; while for $\omega\rightarrow\infty$, it returns to the Schwarzschild solution. The mass function is given by
\begin{align}
\label{m2}
\tilde{m}=\mathcal{M}-\frac{e^{-\frac{r\,\omega\,l^{2}}{\mathcal{M}\,\eta^{2}}}}{2\,\mathcal{M}}\left(\frac{r^{2}\,l^{4}\,\omega^{2}}{\eta^{4}}+\frac{2\,r\,\mathcal{M}\,l^{2}\,\omega}{\eta^{2}}+2\,\mathcal{M}^{2}\right),
\end{align}
which for the case of $r\rightarrow0$ can
be simplified to
\begin{align}
\label{m20}
\tilde{m}=\frac{r^{3}\,l^{6}\,\omega^{3}}{6\,\mathcal{M}^{2}\,\eta^{6}}+\frac{r^{4}\,l^{8}\,\omega^{4}}{12\,\mathcal{M}^{3}\,\eta^{8}}+\mathcal{O}(r^{5}).
\end{align}
This shows that GD deformation acts as a mathematical operation on the seed Schwarzschild metric \eqref{seedSsolution}. Its utility lies in enabling the efficient construction of new BH solutions by imposing desired physical characteristics on the source, rather than through direct physical modification of the seed metric itself.

\begin{figure}[htbp]
    \centering
    
    \begin{minipage}{0.45\textwidth}
        \centering
        \includegraphics[width=\linewidth]{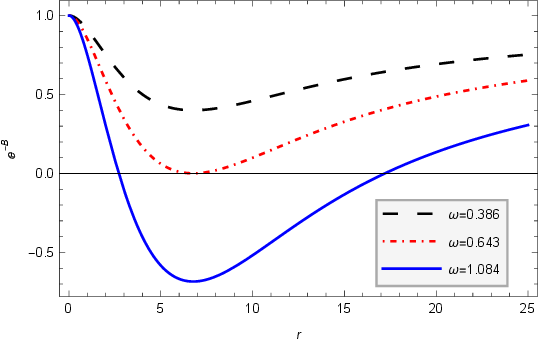}
    \end{minipage}
    \hfill
    \begin{minipage}{0.45\textwidth}
        \centering
        \includegraphics[width=\linewidth]{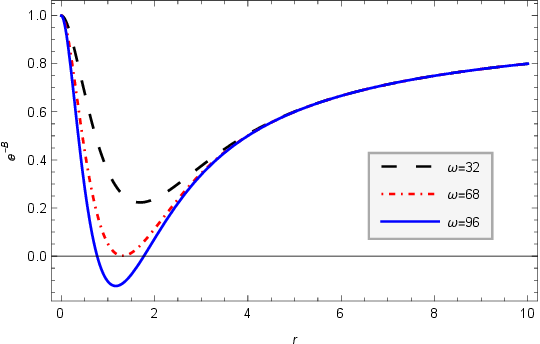}
    \end{minipage}
    
    \vspace{1em} 
    
    \begin{minipage}{0.45\textwidth}
        \centering
        \includegraphics[width=\linewidth]{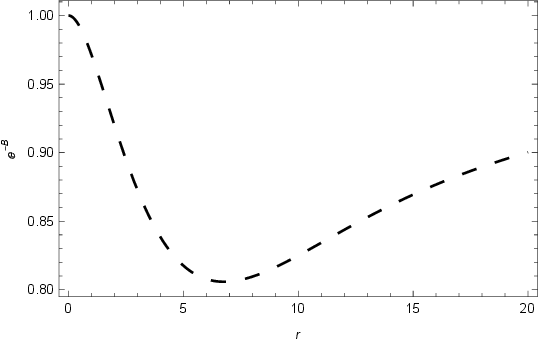}
    \end{minipage}
    \hfill
    \begin{minipage}{0.45\textwidth}
        \centering
        \includegraphics[width=\linewidth]{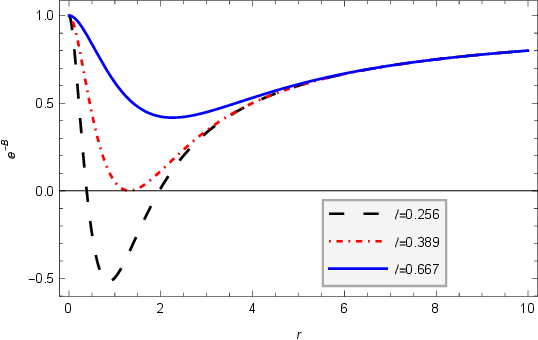}
    \end{minipage}
    
    \caption{The structure of the spherically symmetric metric function \eqref{9g}: no horizons, one horizon, or two horizons if taking $l=2$ and $\eta=1$ (the top left panel), or $\mathcal{M}=1$ and $\eta=2$ (the top right panel), or $\mathcal{M}=1$ and $\omega=2$ (the bottom right panel); and there is no horizon for $\mathcal{M}=1$ and $l=2$ (the bottom left panel)}  
    \label{dingleta1}
\end{figure}

According to equation $e^{-B(r_{\rm h})}=0$, possible horizons can be found. Numerical analysis of this equation shows that there may exist no horizon, one horizon, or two horizons if the non-fixed parameters are greater than, equal to, or less than an extremal value, see the top left, the top right, and the bottom right panel in Figure \ref{dingleta1}. However, there is no horizon for any nonzero $\omega$ (or $\eta$) if we fix the values of $\mathcal{M}$ and $l$, as shown in the bottom left panel of Figure \ref{dingleta1}.

The curvature scalar we find is
\begin{align}
\label{2R}
R=e^{-\frac{r\,\omega\,l^{2}}{\mathcal{M}\,\eta^{2}}}\left(4\,\mathcal{M}\,\eta^{2}-r\,l^{2}\,\omega\right)\frac{l^{6}\,\omega^{3}}{\mathcal{M}^{3}\,\eta^{8}}.
\end{align}
and the Ricci squared is given by
\begin{align}
\label{2RR}
R_{\mu\nu}R^{\mu\nu}=e^{-\frac{2\,r\,\omega\,l^{2}}{\mathcal{M}\,\eta^{2}}}\left(8\,\mathcal{M}^{2}\,\eta^{4}-4\,r\,\mathcal{M}\,\eta^{2}\,l^{2}\,\omega+r^{2}\,l^{4}\,\omega^{2}\right)\frac{l^{12}\,\omega^{6}}{2\,\mathcal{M}^{6}\,\eta^{16}}.
\end{align}
The full expression of the Kretschmann scalar is sufficiently cumbersome, it approximately behaves as for $r\rightarrow0$
\begin{align}
\label{2RRRR}
R_{\mu\nu\rho\sigma}R^{\mu\nu\rho\sigma}\simeq\frac{8\,l^{12}\,\omega^{6}}{3\,\mathcal{M}^{4}\,\eta^{12}}+\frac{20\,r\,l^{14}\,\omega^{7}}{3\,\mathcal{M}^{5}\,\eta^{14}}+\frac{97\,r^{2}\,l^{16}\,\omega^{8}}{12\,\mathcal{M}^{6}\,\eta^{16}}-\frac{15\,r^{3}\,l^{18}\,\omega^{9}}{2\,\mathcal{M}^{7}\,\eta^{18}}+\mathcal{O}(r^{4}),
\end{align}
which shows that the Kretschmann scalar does not diverge when $r\rightarrow0$, meaning that the solution has no curvature singularity.

\begin{figure}[htbp]
    \centering
    
    \begin{minipage}{0.45\textwidth}
        \centering
        \includegraphics[width=\linewidth]{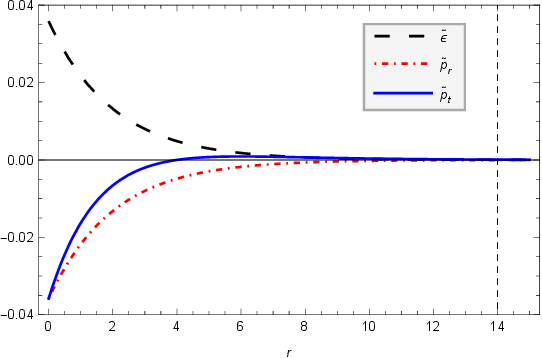}
    \end{minipage}
    \hfill
    \begin{minipage}{0.45\textwidth}
        \centering
        \includegraphics[width=\linewidth]{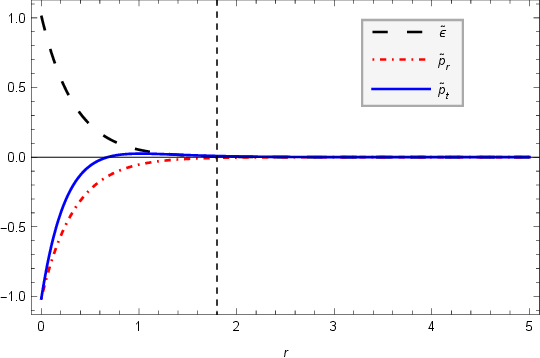}
    \end{minipage}
    
    \vspace{1em} 
    
    \begin{minipage}{0.45\textwidth}
        \centering
        \includegraphics[width=\linewidth]{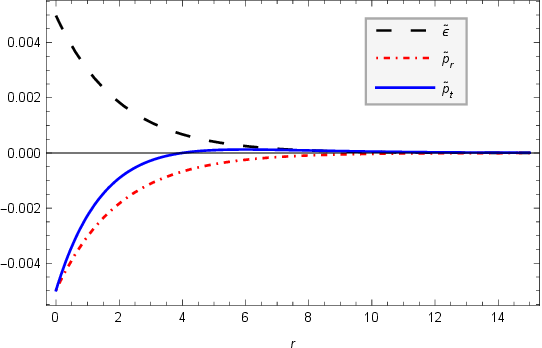}
    \end{minipage}
    \hfill
    \begin{minipage}{0.45\textwidth}
        \centering
        \includegraphics[width=\linewidth]{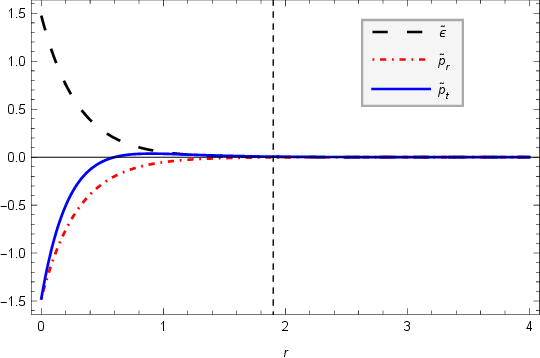}
    \end{minipage}
    
    \caption{The source terms $\tilde{\epsilon}$, $\tilde{p}_{r}$, $\tilde{p}_{t}$ in the sphericity metric case for $l=2$, $\eta=1$ and $\omega=0.9$ (the top left panel); for $\mathcal{M}=1$, $\eta=2$ and $\omega=102$ (the top right panel); for $\mathcal{M}=1$ and $l=2$ (the bottom left panel); and for $\mathcal{M}=1$, $\omega=2$ and $l=0.3$ (the bottom right panel). The vertical dashed lines represent the event horizons, which are located, respectively, at $r_{\rm h}\sim14$, $r_{\rm h}\sim1.8$, and $r_{\rm h}\sim1.9$}
    \label{dingleta3}
\end{figure}

Finally, the effective source giving rise to the metric function \eqref{9g} comprises the energy density defined in Eq. \eqref{C2kE} and the effective tangential pressure
\begin{align}
\label{P02}
\mathcal{P}_{\theta}=e^{-\frac{r\,\omega\,l^{2}}{\mathcal{M}\,\eta^{2}}}\frac{\omega^{3}\,l^{6}}{2\,\kappa\,\mathcal{M}^{3}\,\eta^{8}}\left(r\,l^{2}\,\omega-2\,\mathcal{M}\,\eta^{2}\right).
\end{align}
Together with Eq. \eqref{C2kE}, we further obtain
\begin{align}
\label{EP02}
\mathcal{E}+\mathcal{P}_{\theta}=e^{-\frac{r\,\omega\,l^{2}}{\mathcal{M}\,\eta^{2}}}\frac{r\,l^{8}\,\omega^{4}}{2\,\mathcal{M}^{3}\,\eta^{8}\,\kappa}
\end{align}
As shown in Figure \ref{dingleta3}, the WEC holds within the region $r\ge0$. And the vacuum is rapidly approached beyond the event horizon.
\begin{align}
\label{Prprime2}
\mathcal{P}_{r}'=e^{-\frac{r\,\omega\,l^{2}}{\mathcal{M}\,\eta^{2}}}\frac{l^{8}\,\omega^{4}}{\mathcal{M}^{3}\,\eta^{8}\,\kappa}\ge0.
\end{align}
Equation \eqref{Prprime} indicates that the central pull from a positive radial pressure gradient is offset by the gravitational repulsion induced by the pressure anisotropy $\Pi$.

\subsubsection{Axially symmetric case}
To construct the rotating generalization of the metric function \eqref{9g}, we employ the framework outlined in \cite{Contreras:2021yxe}.
This corresponds to analyzing the general Kerr-Schild metric in Boyer-Lindquist coordinates, that is, the Gurses-Gursey metric \cite{Gurses:1975vu}.
\begin{align}
\label{6g}
ds^{2}=-\bigg[1-\frac{2\,r\,\tilde{m}(r)}{\rho^{2}}\bigg]dt^{2}-\frac{4 \,a\,r\,\tilde{m}(r)\sin^{2}\theta}{\rho^{2}}\,dtd\phi+\frac{\rho^{2}}{\Delta}\,dr^{2}+\rho^{2}\,d\theta^{2}+\frac{\Sigma\,\sin^{2}\theta}{\rho^{2}}\,d\phi^{2},
\end{align}
with
\begin{align}
\rho^{2}&=r^{2}+a^{2}\cos^{2}\theta,                             \label{rho2}    \\
a&=J/\mathcal{M},                                                \label{a}     \\
\Delta&=r^{2}-2\,r\,\tilde{m}(r)+a^{2},                          \label{Delta}   \\
\Sigma&=(r^{2}+a^{2})^{2}-\Delta\,a^{2}\sin^{2}\theta,           \label{Sigma}
\end{align}
where $\tilde{m}$ given by Eq. \eqref{m2} is the mass function of the reference spherically symmetric metric, $\mathcal{M}=\tilde{m}(r\rightarrow\infty)$ is the total mass of the system, and $J$ is the angular momentum. Obviously, we obtain the rotating version of the metric \eqref{9g} without resorting to the Newman-Janis algorithm, which can be reduced to the Kerr solution for $\tilde{m}=M$.

The source $\theta_{\mu\nu}$ for the metric \eqref{6g} can be conveniently recast into the following form 
\begin{align}
 \label{tetrad}
\theta^{\mu\nu}=\tilde{\epsilon}\,u^{\mu}\,u^{\nu}+\tilde{p}_{r}\,l^{\mu}\,l^{\nu}+\tilde{p}_{\theta}\,n^{\mu}\,n^{\nu}+\tilde{p}_{\phi}\,m^{\mu}\,m^{\nu},
\end{align}
with respect to the orthonormal tetrad given by \cite{Gurses:1975vu}
\begin{align}
u^{\mu}&=\frac{\Big(r^{2}+a^{2},0,0,a\Big)}{\sqrt{\Delta\,\rho^{2}}},\qquad                          
l^{\mu}=\frac{\sqrt{\Delta}\Big(0,1,0,0\Big)}{\sqrt{\rho^{2}}},                  \nonumber      \\
n^{\mu}&=\frac{\Big(0,0,1,0\Big)}{\sqrt{\rho^{2}}},\qquad                            
m^{\mu}=-\frac{\Big(a\,\sin^{2}\theta,0,0,1\Big)}{\sqrt{\rho^{2}\,\sin\theta}},                 \\
\kappa\,\tilde{\epsilon}&=-\kappa\,\tilde{p}_{r}=\frac{2\,r^{2}}{\rho^{4}}\,\tilde{m}',         \\
\kappa\,\tilde{p}_{\theta}&=\kappa\,\tilde{p}_{\phi}=-\frac{r^{2}}{\rho^{2}}\,\tilde{m}''+\frac{2\big(r^{2}-\rho^{2}\big)}{\rho^{4}}\,\tilde{m}'.
\end{align}
The metric \eqref{6g} features two distinct types of singularities: $\rho=0$, or $\Delta=0$.
The case of $\rho=0$ identifies to the ring singularity of the Kerr solution, located at $\theta=\pi/2$ and $r=0$, which is a physical singularity. For the limit $a\rightarrow0$, it degenerates to a Schwarzschild-like metric.

The curvature scalar of the metric \eqref{6g} is
\begin{align}
\label{KR}
R=\frac{4\,\tilde{m}'+2\,r\,\tilde{m}''}{\rho^{2}}.
\end{align}
for the mass function \eqref{m2}, it reads
\begin{align}
\label{KerrR2}
R=e^{-\frac{r\,\omega\,l^{2}}{\mathcal{M}\,\eta^{2}}}\frac{r^{2}\,l^{6}\,\omega^{3}}{\mathcal{M}^{3}\,\eta^{8}\,\rho^{2}}\left(4\,\mathcal{M}\,\eta^{2}-r\,l^{2}\,\omega\right).
\end{align}
It is obvious that for $\theta=\pi/2$ and $r\sim0$, Eq. \eqref{KerrR2} is regular. The Ricci squared is completely identical to the regular form presented in Eq. \eqref{2RR}, but the Kretschmann scalar is completely identical to the regular form of Eq. \eqref{2RRRR}. The analysis thus establishes that the rotating solution is devoid of physical singularities.

\begin{figure}[htbp]
    \centering
    
    \begin{minipage}{0.45\textwidth}
        \centering
        \includegraphics[width=\linewidth]{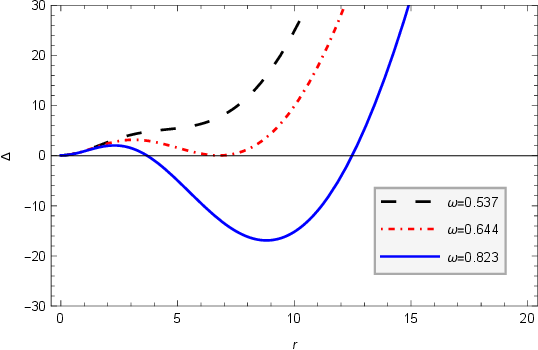}
    \end{minipage}
    \hfill
    \begin{minipage}{0.45\textwidth}
        \centering
        \includegraphics[width=\linewidth]{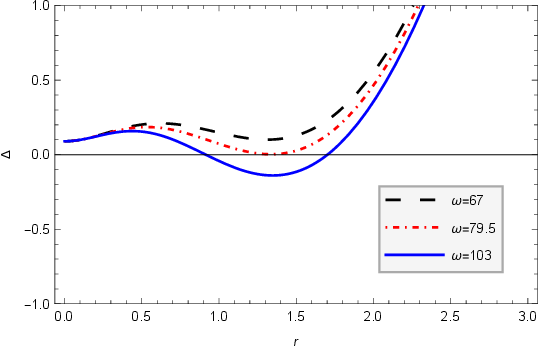}
    \end{minipage}
    
    \vspace{1em} 
    
    \begin{minipage}{0.45\textwidth}
        \centering
        \includegraphics[width=\linewidth]{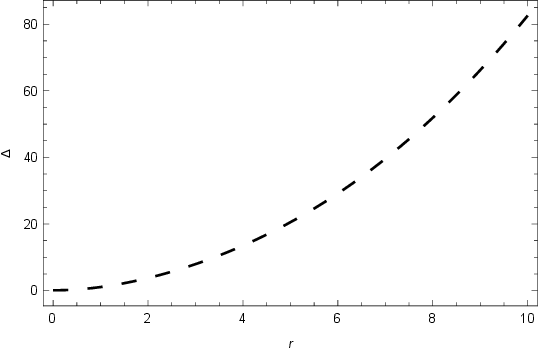}
    \end{minipage}
    \hfill
    \begin{minipage}{0.45\textwidth}
        \centering
        \includegraphics[width=\linewidth]{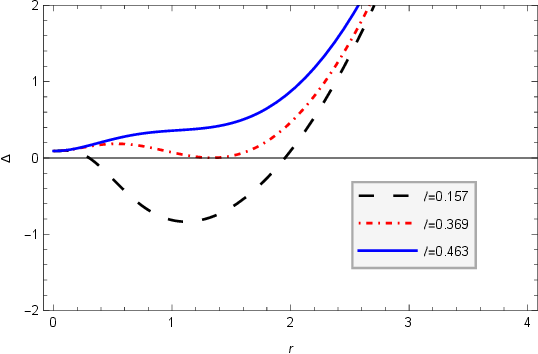}
    \end{minipage}
    
    \caption{Axially symmetric solutions of the metric function \eqref{6g}: There are no horizons, one horizon, and two horizons for $l=2$, $\eta=1$, and $a=0.3$ (the top left panel), or for $\mathcal{M}=1$, $\eta=2$, $a=0.3$ (the top right panel), or for $\mathcal{M}=1$, $\omega=2$, $a=0.3$ (the bottom right panel). There are no horizons for $\mathcal{M}=1$, $l=2$, and $a=0.3$ (the bottom left panel).}
    \label{dingMomega2}
\end{figure}

In general, $\Delta=0$ indicates a coordinate singularity, signaling the presence of a horizon in spacetime.
\begin{align}
\label{0jie2}
\Delta(r_{\rm h})=r_{\rm h}^{2}-2\,r_{\rm h}\,\tilde{m}(r_{\rm h})+a^{2}=0.
\end{align}
Note that, in general, $a\ne0$. As shown in Figure \ref{dingMomega2},
The analysis of Eq. \eqref{0jie2} shows that if fixing the parameters as ($l$, $\eta$, $a$), or ($\mathcal{M}$, $\eta$, $a$), or ($\mathcal{M}$, $\omega$, $a$), there exists an extremal case for $\omega=\omega^*$ or $l=l^*$, with no horizon for $\omega<\omega^*$ or $l>l^*$, and two horizons for $\omega>\omega^*$ or $l<l^*$, which are event and Cauchy horizons, respectively. While if fixing the parameters as ($\mathcal{M}$, $l=2$, $a$), only the extreme configuration is discovered for any value of $\omega$.

\section{conclusions}

It is a fundamental prediction of GR that gravitational collapse inevitably leads to a singularity. To counter this prediction, we introduce a `tensor vacuum', explicitly defined by Eq. \eqref{Prprime}, which functions as a non-collapsing gravitational source, thus providing a mechanism to avoid the singularity.

Within this framework, we construct static and stationary regular BHs parameterized by $\omega$, $\eta$, and $l$. The physical role of $\omega$ is evident from its limiting behavior: as $\omega\rightarrow 0$, both the geometry \eqref{9g} and \eqref{6g} smoothly approach Minkowski spacetime; while as $\omega\rightarrow\infty$ (or $l\rightarrow\infty$, or $\eta\rightarrow 0$), the metric \eqref{9g} reduces to the Schwarzchild geometry, and the metric \eqref{6g} becomes the Kerr spacetime. 

The crucial next steps in future studies can involve exploring the observational implications, stability, and time-dependent formation and evaporation of these solutions. These investigations are of paramount importance for assessing their physical relevance.

\begin{acknowledgments}
This study is supported in part by National Natural Science Foundation of China (Grant No. 12333008).
\end{acknowledgments}

\bibliographystyle{elsarticle-num}
\bibliography{Ref}

\end{document}